\documentclass[]{mn2e}
\usepackage{graphicx}
\newcommand       \Angstrom     {\,{\rm \AA}}

\newcommand       \K            {\,{\rm K}}

\newcommand       \amin         {a_{\rm min}}
\newcommand       \amax         {a_{\rm max}}

\newcommand \mum {\,{\rm \mu m}}

\newcommand \simali {{\sim\,}}
\newcommand       \ppm        {\,{\rm ppm}}
\newcommand       \simlt        {\leq}
\newcommand       \simgt        {\geq}
\newcommand	  \gammabump  {\gamma_{\rm bump}} 
\newcommand	  \xism        {\left[{\rm X/H}\right]_{\rm ism}}
\newcommand	  \xgas        {\left[{\rm X/H}\right]_{\rm gas}}
\newcommand	  \xdust       {\left[{\rm X/H}\right]_{\rm dust}}
\newcommand	  \cism        {\left[{\rm C/H}\right]_{\rm ism}}
\newcommand	  \cgas        {\left[{\rm C/H}\right]_{\rm gas}}
\newcommand	  \cdust       {\left[{\rm C/H}\right]_{\rm dust}}
\newcommand	  \siism        {\left[{\rm Si/H}\right]_{\rm ism}}

\newcommand	  \sidust       {\left[{\rm Si/H}\right]_{\rm dust}}
\newcommand	  \mgism        {\left[{\rm Mg/H}\right]_{\rm ism}}

\newcommand	  \mgdust       {\left[{\rm Mg/H}\right]_{\rm dust}}
\newcommand	  \feism        {\left[{\rm Fe/H}\right]_{\rm ism}}

\newcommand	  \fedust       {\left[{\rm Fe/H}\right]_{\rm dust}}
\newcommand	  \xsun         {\left[{\rm X/H}\right]_\odot}
\newcommand	  \sisun        {\left[{\rm Si/H}\right]_\odot}
\newcommand	  \mgsun        {\left[{\rm Mg/H}\right]_\odot}
\newcommand	  \fesun        {\left[{\rm Fe/H}\right]_\odot}
\newcommand	  \csun        {\left[{\rm C/H}\right]_\odot}

\title[Buckyonions vs. the 2175$\Angstrom$
       Interstellar Extinction Feature]
  {On Buckyonions as an Interstellar Grain Component}
\author[A.~Li et al.]
  {Aigen Li$^{1}$, 
   J.H. Chen$^{1,2}$, 
   M.P. Li$^1$, 
   Q.J. Shi$^{3}$,
   and Y.J. Wang$^2$\thanks{%
   E-mail: lia@missouri.edu}\\
  $^1$Department of Physics \& Astronomy, 
      University of Missouri, Columbia, MO 65211, USA\\
  $^2$Department of Physics, Hunan Normal University,
      Changsha, Hunan 410081, China\\
  $^3$Department of Science and Engineering
      of Shuda College, Hunan Normal University,
      Changsha, Hunan 410081, China\\
      }      

\begin{document}
\date{Received date  / Accepted date }
\pagerange{\pageref{firstpage}--\pageref{lastpage}} \pubyear{2007}

\maketitle

\label{firstpage}
\begin{abstract}
The carrier of the 2175$\Angstrom$ interstellar extinction feature
remains unidentified since its first detection over 40 years ago.
In recent years carbon buckyonions have been proposed as a carrier 
of this feature, based on the close similarity between the electronic 
transition spectra of buckyonions and the 2175$\Angstrom$ interstellar
feature. We examine this hypothesis by modeling the interstellar
extinction with buckyonions as a dust component. It is found that
dust models containing buckyonions (in addition to amorphous 
silicates, polycyclic aromatic hydrocarbon molecules, graphite) 
can closely reproduce the observed interstellar extinction curve. 
To further test this hypothesis, we call for experimental
measurements and/or theoretical calculations of the infrared
vibrational spectra of hydrogenated buckyonions.
By comparing the infrared emission spectra predicted 
for buckyonions vibrationally excited 
by the interstellar radiation 
with the observed emission spectra of the diffuse
interstellar medium, we will be able to derive 
(or place an upper limit on) the abundance of 
interstellar buckyonions.
\end{abstract}


\section{Introduction}
In the interstellar medium (ISM), the strongest spectroscopic 
extinction feature is the 2175$\Angstrom$ bump which is
characterized with a stable peak wavelength and a width
variable from one sightline to another 
(Fitzpatrick \& Massa 2007).
Since Stecher (1965) first detected this ultraviolet (UV) 
extinction feature through rocket observations, 
the origin of this feature and the nature of 
its carrier(s) are still an enigma.
Many candidate materials, including graphite, amorphous carbon,
graphitized (dehydrogenated) hydrogenated amorphous carbon, 
nano-sized hydrogenated amorphous carbon,
quenched carbonaceous composite, coals, 
polycyclic aromatic hydrocarbon (PAH),
and OH$^{-}$ ion in low-coordination sites on or within 
silicate grains have been proposed, while no single one 
is generally accepted 
(see Li \& Greenberg 2003 for a review). 

Recently, Chhowalla et al.\ (2003) measured the 
UV-visible photoabsorption spectra
of carbon buckyonions (BOs)\footnote{%
  Early pioneering experimental studies of
  the UV absorption properties of BOs and their
  association with the 2175$\Angstrom$ interstellar
  extinction feature include those of
  de Heer \& Ugarte (1993), Ugarte (1995), 
  and Wada et al.\ (1999).
  }
composed of spherical concentric fullerene shells
(so far the largest BOs produced in laboratory 
 have $N\sim 100$ shells; Iglesias-Groth et al.\ 2003).
They found that the plasmon-like feature of BOs due to 
a collective excitation of the $\pi$ electrons closely 
fits the 2175$\Angstrom$ interstellar extinction feature. 
More recently, Ruiz et al.\ (2005) theoretically simulated 
the photoabsorption spectra of BOs.\footnote{%
  Early pioneering theoretical studies of
  the UV absorption properties of BOs and their
  association with the 2175$\Angstrom$ interstellar
  extinction feature include those of Wright (1988),
  Henrard et al.\ (1993, 1997) and Lucas et al.\ (1994).
  }
They found that the calculated absorption spectra of BOs
are in close agreement with the experimental data of 
Chhowalla et al.\ (2003) and the observed 2175$\Angstrom$ 
interstellar feature.
That the $\pi$-plasmon absorption band of BOs exhibits 
a stable peak position at $\simali$5.70\,eV 
($\lambda\approx 2175\Angstrom$, $\lambda^{-1}\approx 4.6\mum^{-1}$) 
and a variable bandwidth
(Chhowalla et al.\ 2003, Ruiz et al.\ 2005) suggests that
BOs may be a promising candidate material for the 2175$\Angstrom$
interstellar extinction feature.

Indeed, it is shown by Chhowalla et al.\ (2003) and
Ruiz et al.\ (2005) that the photoabsorption spectra
of BOs {\it alone} very well reproduce 
the entire interstellar extinction curve
at $\lambda^{-1}$\,$\simali$3.2--7.3$\mum^{-1}$.   
This requires $\simali$190\,ppm (parts per million) 
C/H to be locked up in BOs (Ruiz et al.\ 2005).  
However, it is well recognized that in the ISM,
in addition to the 2175$\Angstrom$ extinction carrier,
there must exist other dust components as well --
there must be a population of amorphous silicate
dust, as indicated by the strong, ubiquitous 9.7 and 18$\mum$
interstellar absorption features; there must be a population
of aromatic hydrocarbon dust (presumably PAH molecules), 
as indicated by the distinctive set of ``unidentified'' 
infrared (UIR) emission bands at 3.3, 6.2, 7.7, 8.6, 
and 11.3$\mum$ ubiquitously seen in the ISM; 
there must also exist a population of 
aliphatic hydrocarbon dust,
as indicated by the 3.4$\mum$ C--H absorption feature
which is also ubiquitously seen in the diffuse ISM of
the Milky Way and external galaxies (see Li 2004).

Although buckyonions are able to closely reproduce
the 2175$\Angstrom$ extinction feature, it is not 
clear if dust models with amorphous silicates, PAHs, and 
other carbon dust species (e.g. amorphous carbon, hydrogenated 
amorphous carbon, organic refractory, and graphite) incorporated 
(in addition to BOs) are still capable of fitting the 2175$\Angstrom$ 
extinction feature. One may intuitively expect that the almost 
perfect fit to the $\lambda^{-1}$\,$\simali$3.2--7.3$\mum^{-1}$ 
interstellar extinction would easily be distorted by 
the addition of other dust components.
It is the purpose of this {\it Letter} to examine this issue.
To this end, we consider the extinction of dust models 
consisting of multi-modal grain populations with BOs as
an interstellar grain component.

\begin{figure}
\begin{center}
\includegraphics{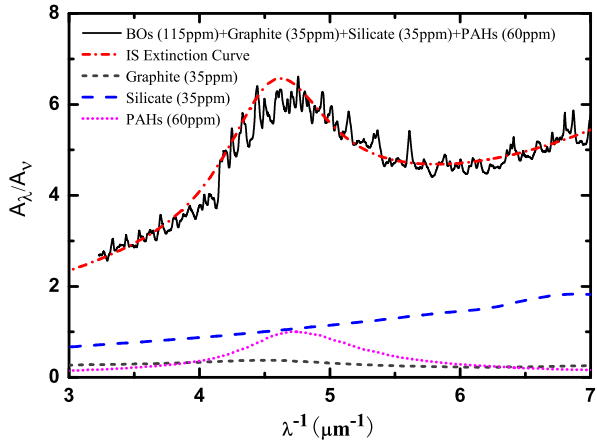}
\end{center}
\caption{\label{fig:mrn_solar}
         Comparison of the interstellar extinction (dot-dashed line)
         with the model extinction (solid line with oscillatory features) 
         obtained by summing up the contributions of buckyonions 
         (C/H\,=\,115\,ppm), PAHs (C/H\,=\,60\,ppm),
         graphite (C/H\,=\,35\,ppm), and
         amorphous silicate (Si/H\,=\,35\,ppm).
         We assume a MRN-type size distribution for         
         the silicate and graphite grains.
         }
\end{figure}


\section{Models}
We consider dust models consisting of amorphous silicate,
graphite, PAHs, and BOs. We take the size distributions 
of silicate and graphite dust to be either that of Mathis, 
Rumpl, \& Nordsieck (1977; hereafter MRN) or that of 
Weingartner \& Draine (2001; hereafter WD).
The MRN size distribution is a simple power-law
$dn/da \sim a^{-3.5}$ with a lower cutoff at
$\amin = 50\Angstrom$ and an upper cutoff at
$\amax = 0.25\mum$ for both silicate and graphite. 
The WD size distribution is more extended.
Unlike the MRN distribution, it smoothly extends from 
$\amin=3.5\Angstrom$ up to $\simali$1$\mum$
for both silicate and graphite.\footnote{%
  The optical properties of graphite are taken to
  be that of PAHs at very small sizes ($a\simlt 25\Angstrom$)
  and that of graphite at radii $a\simgt 100\Angstrom$
  (see Li \& Draine 2001).
  }
We also include a PAH component with a log-normal 
size distribution 
$dn/d\ln a \propto \exp\{-[{\rm ln}(a/a_{\rm o})]^2/(2\sigma^2)\}$
for $a \simgt 3.5\Angstrom$, 
where $a_{\rm o}=3.5\Angstrom$ and $\sigma=0.4$ determine 
the peak location and width of the distribution
(Li \& Draine 2001, Weingartner \& Draine 2001).
For BOs, we do not need to consider size distributions
since the UV-visible photoabsorption spectra are
practically independent of size for BOs with 
the number of nested shells $N>3$
(Chhowalla et al.\ 2003, Ruiz et al.\ 2005).

\begin{figure}
\begin{center}
\includegraphics{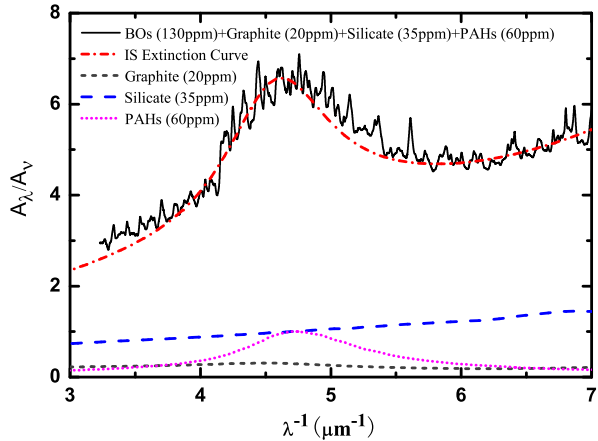}
\end{center}
\caption{\label{fig:wd_solar}
         Same as Figure \ref{fig:mrn_solar} but 
         with a WD-type size distribution assumed for         
         the silicate and graphite grains.        
         }
\end{figure}


The quantity of dust is taken to be consistent with 
the interstellar abundance constraints: the dust-forming 
elements (C, O, N, Si, Mg, Fe), if not seen in the gas phase, 
must have been depleted into dust. This requires an accurate 
knowledge of the interstellar reference abundance 
(i.e. the total abundance of an element both in gas and in dust). 

Let $\xism$ be the interstellar reference abundance 
of element X relative to H,
$\xgas$ be the gas-phase abundance of X (relative to H).
The abundance of X (relative to H) in dust is
$\xdust = \xism - \xgas$. 
We assume that all cosmically available Si, Mg, and Fe 
are locked up in amorphous silicate dust, 
i.e. $\sidust = \siism$, $\mgdust = \mgism$, 
and $\fedust = \feism$ 
(see Savage \& Sembach 1996).
We take $\cgas = 140\ppm$ (Cardelli et al.\ 1996).

We assume the interstellar reference abundance 
to be solar, i.e. $\xism = \xsun$
($\cism = \csun \approx 361\ppm$,
$\siism = \sisun \approx 35\ppm$,
$\mgism = \mgsun \approx 36\ppm$,
$\feism = \fesun \approx 30\ppm$).\footnote{%
   The published solar abundances have undergone 
   major changes over the years and are still subject 
   to major systematic uncertainties, 
   as demonstrated in Table 1 of Li (2005).
   We take the average values of the widely used 
   solar abundances compiled by 
   Grevesse \& Sauval (1998) and Holweger (2001).
   }

With $\cgas = 140\ppm$ in the gas phase,
we have $\cdust \approx 221\ppm$ left for carbon dust:
buckyonions, PAHs, and graphite.
The ``UIR'' bands require C/H\,$\approx$\,60ppm
to be in PAHs (Li \& Draine 2001).
So we have C/H\,=\,161\,ppm for buckyonions and graphite.
For silicate dust we assume a stoichiometric composition 
of MgFeSiO$_4$ and Si/H\,=\,Mg/H\,=\,Fe/H\,=\,35\,ppm.

With these parameters specified, we just vary the amounts
of C (relative to H) depleted in BOs and graphite to fit
the interstellar extinction curve. 
As shown in Figure \ref{fig:mrn_solar}, 
with C/H\,=\,35\,ppm in graphite
and C/H\,=\,115\,ppm in BOs,
the dust model consisting of multi-modal grain populations 
and with a MRN-type size distribution
closely fits the interstellar extinction curve
at $\lambda^{-1}$\,$\simali$\,3.2--7$\mum^{-1}$
(at present the photo-absorption spectra of buckyonions
are available only in this wavelength range).
This model requires C/H\,=\,210\,ppm,
smaller than the total available 
$\cdust = 221\ppm$.
The extinction cross sections of amorphous silicate
and graphite are calculated from Mie theory [using
the dielectric functions of Draine \& Lee (1994)
and assuming a spherical shape for these grains].
The absorption spectra of PAHs and BOs are taken
respectively from Draine \& Li (2007) and Ruiz et al.\ (2005).
The latter was an average of eight BOs 
for which the number of nested shells
$N = 3,4,...10$ (see Ruiz et al.\ 2005).
Note that the UV-visible absorption spectra
of BOs are essentially independent of $N$ (Ruiz et al.\ 2005).

Similarly, we have considered a WD-type size distribution
for the silicate and graphite dust components.
As shown in Figure \ref{fig:wd_solar}, 
with C/H\,=\,20\,ppm in graphite
and C/H\,=\,130\,ppm in BOs,
the multi-component grain model 
provides a reasonably close match to the interstellar
extinction curve. The total required carbon abundance
is also C/H\,=\,210\,ppm,
smaller than the total available 
$\cdust = 221\ppm$.

The oscillatory features in the BO model spectra
arise from the electronic transitions of BOs.
As stated in Ruiz et al.\ (2005),
not all the oscillations in the photoabsorption spectra
of BOs are reliable, but the general features such as
the average width and peak position are quite
robust and reliable.
If BOs are indeed present in the ISM,
we would expect a mixture of BOs of various sizes
and various degrees of hydrogenation, ions, and radicals
of which the oscillatory features occur at different
wavelengths; therefore, as a collective effect,
the strong indentations seen both experimentally 
and theoretically in the photoabsorption spectra 
of individual BOs will likely be smoothed out and
will not show up in astronomical spectra.
 
\section{Discussion}
It is encouraging that the dust models consisting of
multi-modal grain populations (including BOs) fit the
interstellar extinction curve (including the 2175$\Angstrom$
extinction feature) very well while satisfying the interstellar
abundance constraints. The fit to the 2175$\Angstrom$
feature is mainly affected by the silicate dust component.
The smaller amount of silicate dust is included in the model,
the better is the fit. We would achieve a closer match to
the 2175$\Angstrom$ extinction if we take the interstellar
Si abundance to be subsolar (Snow \& Witt 1996)
like that of B stars $\sidust=18\ppm$ (Sofia \& Meyer 2001)
while having C/H\,=\,140\,ppm in BOs and C/H\,=\,10\,ppm 
in graphite.\footnote{%
  If both C and Si are subsolar 
  (say, 2/3 of that of solar, see Snow \& Witt 1996),
  we will not be able to fit the interstellar extinction
  (also see Li 2005).
  }

BOs are a promising candidate material for the 2175$\Angstrom$
interstellar extinction feature: (1) their $\pi$-plasmon 
absorption profile closely resembles the 2175$\Angstrom$
extinction feature and is stable in peak wavelength position 
while varies in width;\footnote{%
  Ruiz et al.\ (2005)'s theoretically simulated photoabsorption
  spectra of BOs showed that the width of the 2175$\Angstrom$ feature
  $\gammabump$ varies with size for small BOs, 
  however, the width becomes independent of size when the number 
  of shells $N\simgt 6$ 
  ($\gammabump \approx 1.03\mum^{-1}$; see their Fig.\,4).
  For BOs to explain the 2175$\Angstrom$ interstellar
  features broader than $\gammabump = 1.03\mum^{-1}$
  (some sightlines have $\gammabump$ up to $\simali$1.2$\mum^{-1}$),
  clustering of individual BOs  
  (de Heer \& Ugarte 1993, Rouleau et al.\ 1997), 
  imperfect growth of BOs (e.g. mixing with amorphous carbon
  impurities; see Chhowalla et al.\ 2003),
  and coating BOs with a layer of PAHs (Mathis 1994)
  may play an important role.
  } 
(2) the almost perfect fit
to the interstellar feature is not significantly 
distorted by the inclusion of other dust components
(e.g. silicate, PAHs, graphite) which are required
to account for other interstellar phenomena 
(e.g. the 9.7, 18$\mum$ absorption features,
the ``UIR'' bands); and (3) BOs are highly stable molecules 
that they can survive under intense UV radiation and are 
highly resistant to destruction by collisions.

BOs can be formed by electron beam irradiation of 
carbon soot (Ugarte 1992) and by heat treatment 
(annealing at temperatures $T\simgt 700\,^{\circ}$C) 
and electron beam irradiation of nanodiamond 
(Kuznetsov et al.\ 1994, Tomita et al.\ 2002).
The generation of BOs by annealing nanodiamonds
is astrophysically relevant.
Presolar nanodiamonds are identified in
primitive carbonaceous meteorites based on 
their isotopic anomalies (Lewis et al.\ 1987).
The possible existence of nanodiamonds in dense 
clouds and circumstellar dust disks or envelopes
have also been suggested 
(Allamandola et al.\ 1992, 
Guillois, Ledoux, \& Reynaud 1999, 
van Kerckhoven, Tielens, \& Waelkens 2002).
The generation of BOs through annealing nanodiamonds
can occur in the ISM where nanodiamonds are stochastically
heated to temperatures as high as $\simali$1000$\K$
by the interstellar radiation field
(Jones \& d'Hendecourt 2000).
Indeed, BOs have been found in meteorites 
with anomalous isotopic compositions
(Smith \& Buseck 1981, Bernatowicz et al.\ 1996, Harris et al.\ 2000).

Recently, BOs have also been invoked to explain
other interstellar phenomena: the mysterious diffuse 
interstellar bands (DIBs) resulting from the electronic
transitions of BOs (Iglesias-Groth 2004, 2007),\footnote{%
  To date, $>$300 DIBs have been detected in the Galactic
  and extragalactic ISM. These mysterious optical-to-near-IR
  absorption lines (broader than the narrow lines
  from gas atoms, ions, and small molecules) 
  still remain unidentified.
  Fullerenes are known to have strong structural resonances 
  at this wavelength range 
  (Dresselhaus et al.\ 1996, Iglesias-Groth 2004).
  Iglesias-Groth (2007) argued that BOs may be
  responsible for the strongest optical DIB at 4430$\Angstrom$ 
  and possibly also the 6177$\Angstrom$ and 6284$\Angstrom$ DIBs.
  }
the 10--100\,GHz Galactic anomalous microwave emission
resulting from the electric dipole emission of rotationally
excited BOs (Iglesias-Groth 2005, 2006), 
and the broad feature around 100$\mum$ 
of the diffuse emission from two active star-forming regions 
(the Carina Nebula and Sharpless 171)
resulting from the small particle surface resonance 
(Onaka \& Okada 2003). 

Experimentally, all BOs are found to belong to 
the C$_{60N^2}$ icosahedral family
with $N=1,2,...$ (i.e. their shells are consecutive 
elements of the icosahedral C$_{60N^2}$ fullerene family,
with C$_{60}$ being the smallest shell\footnote{%
  Kroto et al.\ (1985) first proposed that C$_{60}$
  could be present in the ISM with a considerable
  quantity. This molecule and its related species 
  were later proposed as the carriers of 
  the 2175$\Angstrom$ extinction hump, 
  the DIBs, the ``UIR'' bands, 
  and the extended red emission 
  (see Webster 1991, 1992, 1993a,b).
  C$_{60}$ and C$_{70}$ are unlikely a major contributor
  to the 2175$\Angstrom$ extinction feature since they
  have a characteristic doublet absorption 
  in this wavelength region. 
  Foing \& Ehrenfreund (1994) attributed the two
  DIBs at 9577$\Angstrom$ and 9632$\Angstrom$ to C$_{60}^{+}$.
  However, attempts to search for these molecules in
  the UV and IR were unsuccessful (Snow \& Seab 1989; 
  Somerville \& Bellis 1989; Moutou et al.\ 1999; Herbig 2000).
  These molecules are now estimated to consume at most 
  $<$0.7$\ppm$ carbon (Moutou et al.\ 1999). 
  Therefore, C$_{60}$ is at most a minor 
  component of the interstellar dust family. 
  }
and the intershell distance being very close to 
the interplanar separation in graphite, often taken 
to be 3.55$\Angstrom$, the C$_{60}$ radius;
Yoshida \& 1993, Ruiz et al.\ 2004).
So far,  all the experimentally generated and analyzed BOs
are in the nanometer size range
($\simali$3--15\,nm in diameter; 
see de Heer \& Ugarte 1993,
Cabioc'h et al.\ 1997,
Chhowalla et al.\ 2003).


The fullerenes produced in laboratory are usually accompanied 
by tubular and other non-spherical graphite-like structures of 
which the theoretical UV spectra exhibit some similarity to 
that of BOs. In the ISM, tubular graphite and PAHs can
be catalytically formed at $T\simgt 1000\K$ on Fe nanoparticles
and are probably the carrier of some of the DIBs 
(Zhou et al.\ 2006). They may also contribute to 
the 2175$\Angstrom$ interstellar extinction feature. 

Finally, we should note that a powerful test of the BOs hypothesis 
would be in the IR. Because of their small sizes (and therefore small 
heat capacities), in the ISM, BOs will be stochastically heated by 
single UV photons (Draine \& Li 2001). With their surface shell 
hydrogenated, they will emit in the near- to mid-IR through 
their characteristic C--H stretching and bending bands, 
and C--C stretching bands. The detection (or non-detection) of 
these bands will allow us to derive (or place an upper limit on) 
the abundance of BOs. Note that the BO models require an appreciable
amount of C/H to be in BOs ($>110\ppm$), nearly twice as much as PAHs.
Unfortunately, little is known about 
the positions and strengths of these vibrational bands. 
We call for urgent experimental measurements and/or theoretical 
calculations of the IR vibrational spectra of hydrogenated BOs.
Future far-IR/submm space missions (e.g. Herschel)
will also be useful for studying their far-IR vibrational
spectra (see Iglesias-Groth \& Bret\'on 2000). 
Moreover, it would be interesting to
see if the silicate-graphite-PAH-BO multi-component dust model 
is able to reproduce the $\simali$2--3000$\mum$
overall IR emission of the Galactic ISM.

\section*{Acknowledgments}
We thank J.M. Gomez Llorente for sending us 
the photoabsorption spectrum of BOs. 
We thank the anonymous referee for 
his/her helpful comments. 
We are supported in part
by NASA/HST Theory Programs
and NSF grant AST 07-07866.

\label{lastpage}

\end{document}